\documentclass[aps,prb,reprint,superscriptaddress,amsmath,amssymb,showpacs,longbibliography]{revtex4-2}
\usepackage[utf8]{inputenc}
\usepackage{graphicx,amsmath,amssymb,times,float,color,csquotes}

\date{22 November 2022}

\begin{document}

\title{Many-body localization and the area law in two dimensions}

\author{K.~S.~C.~Decker}
\affiliation{Technische Universit\"at Braunschweig, Institut f\"ur Mathematische Physik, Mendelssohnstraße 3, 38106 Braunschweig, Germany}

\author{D.~M.~Kennes}
\affiliation{Institut f\"ur Theorie der Statistischen Physik, RWTH Aachen University and JARA-Fundamentals of Future Information Technology, 52056 Aachen, Germany}
\affiliation{Max Planck Institute for the Structure and Dynamics of Matter, Center for Free-Electron Laser Science, 22761 Hamburg, Germany}

\author{C.~Karrasch}
\affiliation{Technische Universit\"at Braunschweig, Institut f\"ur Mathematische Physik, Mendelssohnstraße 3, 38106 Braunschweig, Germany}

\begin{abstract}
We study the high-energy phase diagram of a two-dimensional spin-$\frac{1}{2}$ Heisenberg model on a square lattice in the presence of either quenched or quasiperiodic disorder. The use of large-scale tensor network numerics allows us to compute the bipartite entanglement entropy for systems of up to $60\times7$ lattice sites. We provide evidence for the existence of a many-body localized regime for large disorder strength that features an area law in excited states and that violates the eigenstate thermalization hypothesis. From a finite-size analysis, we determine an estimate for the critical disorder strength where the transition to the ergodic regime occurs in the quenched case. 
\end{abstract}

\maketitle

\section{Introduction}

Paradigmatic to thermodynamics is the assumption that almost any initial state of a generic closed system will evolve into a thermal state governed by the expectation values of a few macroscopic quantities such as energy or particle number \cite{Polkovnikov2011,Gogolin2016}. Many-body localization (MBL) provides a hallmark exception to this paradigm and generalizes the well-understood phenomena of Anderson localization \cite{Anderson1958} to the interacting realm. Even in the presence of interactions, too much information of the initial state is retained, which hinders equilibration and the description by a thermal ensemble \cite{Gornyi2005,Basko2006,Pal2010,Abanin2019}. Since the pioneering works of \cite{Gornyi2005,Basko2006}, many-body localization has raised tremendous attention as an increasing number of astonishing properties of this phase were uncovered. These include a logarithmic growth of entanglement in quenched systems \cite{Bardarson2012,Serbyn2013,Znidaric2008}, locality of information spread celebrated for potential applications in quantum information sciences \cite{Banuls2017,Friesdorf2015}, unusual transport properties \cite{Luitz2017,Agarwal2017,BarLev2017}, an area law \cite{Eisert2010} scaling of the entanglement in the {\it excited states} \cite{Bauer2013}, and the $\ell$-bit picture \cite{Serbyn2013}. These findings were substantiated in one spatial dimension by extensive numerical studies that employed mainly exact diagonalization or tensor network approaches  \cite{Abanin2019}. However, the fate of MBL in the thermodynamic limit was recently questioned as discussed below \cite{PhysRevB.100.104204,PhysRevE.102.062144,PhysRevB.103.024203,PhysRevE.104.054105,PhysRevLett.127.230603,ABANIN2021168415,PhysRevLett.124.186601,Panda_2020,PhysRevB.102.100202,PhysRevB.105.144203,PhysRevB.105.174205,PhysRevB.95.155129,PhysRevLett.121.140601,PhysRevLett.119.150602,PhysRevB.99.195145,PhysRevResearch.2.033262,PhysRevB.100.115136,PhysRevB.106.L020202,https://doi.org/10.48550/arxiv.2012.15270}.

These extensive theoretical efforts were soon complemented by experimental approaches from the field of cold quantum gases \cite{Schreiber2015,Smith2016} or photons \cite{Roushan2017}, which allow an accurate simulation of the thermalization dynamics of quantum many-body systems, or the lack thereof in the case of MBL. Experimental setups are less restrictive in terms of dimensionality, size, time scales, or entanglement growth compared to numerical approaches. They are, however, restricted to studying the evolution of a few initial states, and while quench dynamics can be directly simulated, access to the many-body spectrum -- showing many of the prototypical hallmarks of MBL -- remains elusive using this experimental route. In addition, experimental setups are not concentrating on quenched but on quasiperiodic disorder, which is much more convenient to implement. 

\begin{figure}[b]
\centering
\includegraphics[width=0.9\linewidth]{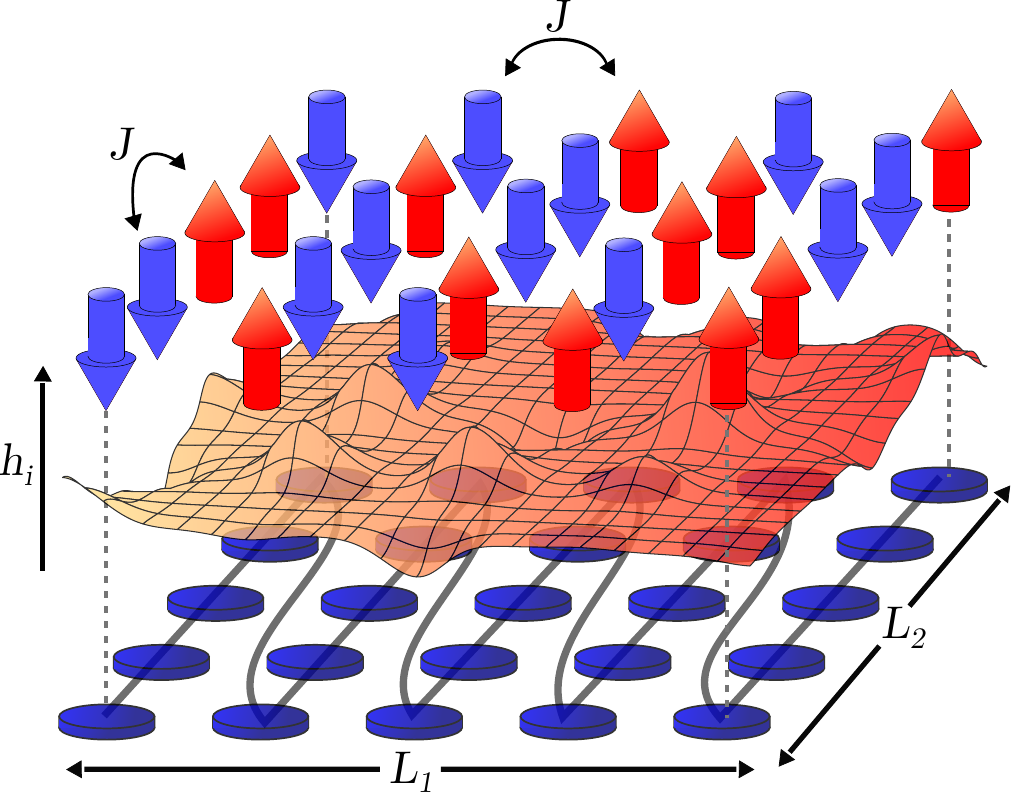}
\caption{Sketch of the system studied in this Letter. Quantum-mechanical spin-$\frac{1}{2}$ degrees of freedom $\hat{\mathbf{S}}_{i}$ are placed on a two-dimensional square lattice of horizontal and vertical dimension $L_1$ and $L_2$, respectively. They interact via a Heisenberg exchange term [see Eq.~(\ref{eq:h})]. The strength of the on-site magnetic field $h_i$ fluctuates (\enquote{random potential landscape}) or follows a quasiperiodic distribution. The snakelike configuration of the tensor network is indicated at the bottom. The bipartite entanglement entropy in Figs.~\ref{fig:ent}, \ref{fig:ent2}, and \ref{fig:quasi} is calculated for vertical cuts.}
\label{fig:system}
\end{figure}

Taken together, theoretical and experimental advances have significantly deepened our understanding of MBL in one spatial dimension. First forays into the two-dimensional regime have been undertaken on the level of the quench dynamics, both experimentally \cite{Choi2016,Bordia2017} and from the numerical point of view \cite{Kennes2018,Hubig2019,DeTomasi2019,Kshetrimayum2020,Doggen2020}, each with its distinct restrictions. Although some evidence for MBL in two-dimensional systems \cite{Wahl2019,Chertkov2021,Tang2021} has recently been accumulated, the state of affairs is much less clear, stimulating a topical debate about whether MBL might ultimately be unstable in two-dimensions \cite{DeRoeck2017a,DeRoeck2017b}.

\begin{figure*}[t]
\centering
\includegraphics[width=0.9\linewidth]{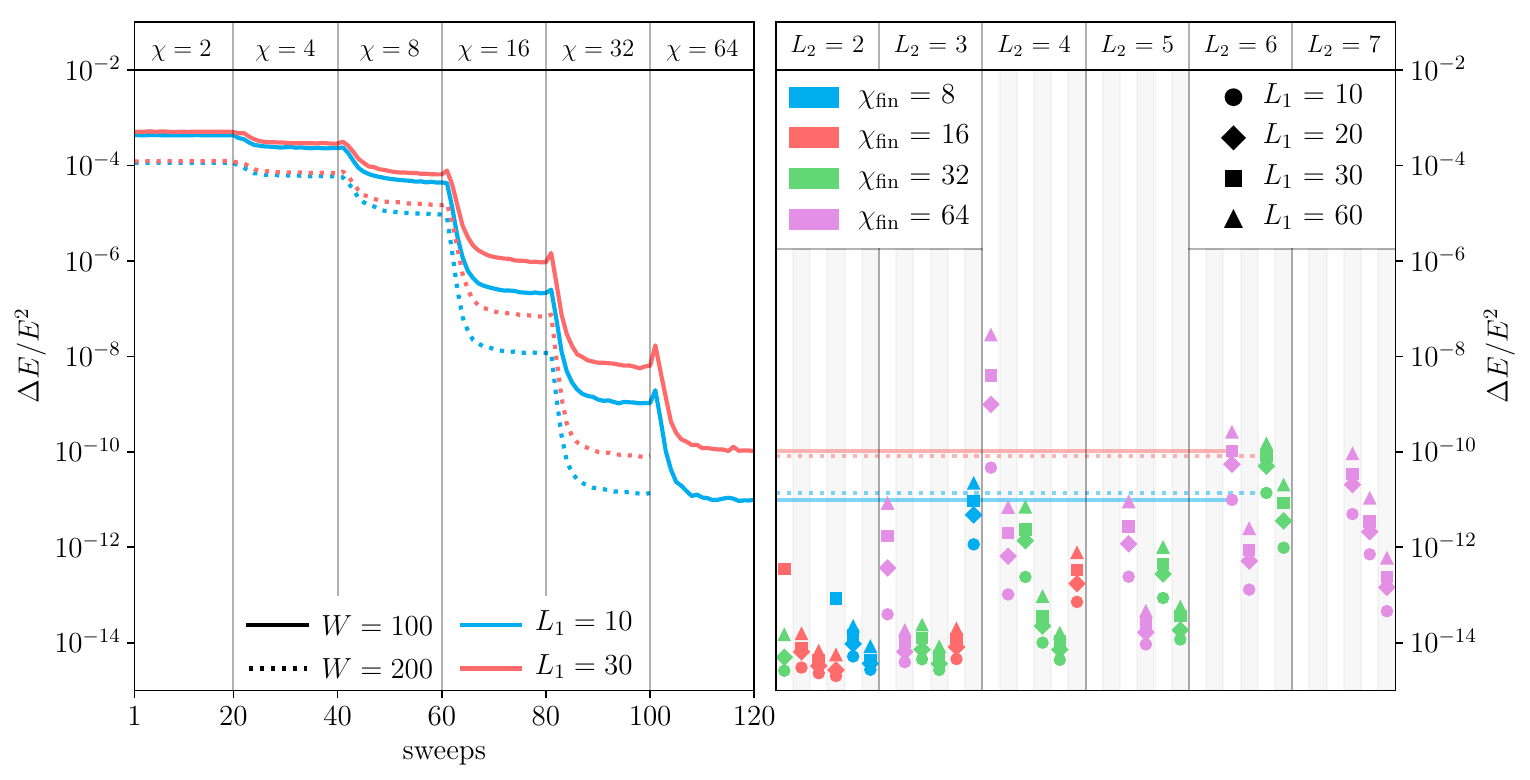}
\caption{\textit{Left panel:} Rescaled energy variance $\Delta E / E^2$ of high-energy eigenstates as a function of the DMRG-X sweeps for a system of length $L_1\in\{10,30\}$, width $L_2=6$, and open boundary conditions for two different strengths $W\in\{100,200\}$ of a quenched disorder potential. The local bond dimension $\chi$ is successively increased until $\Delta E / E^2$ has sufficiently converged. By exploiting $U(1)$-symmetries, we can work with values as large as $\chi=64$. \textit{Right panel:} Scatter plot of the energy variance at the end of the corresponding DMRG-X sweeping procedure. The individual data points correspond to all parameter sets $\{L_1,L_2,W\}$ shown in Fig.~\ref{fig:ent}. For fixed $L_2$, the disorder strength $W$ increases from left to right. The different colors indicate the \textit{final} value of the local bond dimension for the corresponding parameters. The solid and dashed lines illustrate which data points correspond to the four parameter sets shown in the left panel.}
    \label{fig:dmrg}
\end{figure*}

In this Letter, we employ an accurate numerical approach to establish evidence for many-body localization in spin-$\frac{1}{2}$ Heisenberg models on a square lattice with both quenched or quasiperiodic disorder potentials. In particular, we show that highly excited eigenstates of the system obey an area law scaling. From the prefactor of this scaling law, we estimate the critical disorder strength where the transition into an ergodic phase occurs in the case of quenched disorder.

Many studies of MBL in one dimension are based on an exact diagonalization of small systems \cite{Pal2010,Luitz2015}. It was recently questioned if the MBL phase indeed exists in the thermodynamic limit \cite{PhysRevB.100.104204,PhysRevE.102.062144,PhysRevB.103.024203,PhysRevE.104.054105,PhysRevLett.127.230603}, but a conclusive picture has yet to emerge \cite{ABANIN2021168415,PhysRevLett.124.186601,Panda_2020,PhysRevB.102.100202,PhysRevB.105.144203,PhysRevB.105.174205}. Quantum avalanches potentially provide a new perspective \cite{PhysRevB.95.155129,PhysRevLett.121.140601,PhysRevLett.119.150602,PhysRevB.99.195145,PhysRevResearch.2.033262,PhysRevB.100.115136,PhysRevB.106.L020202,https://doi.org/10.48550/arxiv.2012.15270}. In this Letter, we address the finite-size MBL regimes in two dimensions in analogy to the one-dimensional case. Our aim is to provide large-scale, state-of-the-art tensor network data. We refrain from commenting on the fate of the MBL phase in the thermodynamic limit.

\section{Model and Method}

We study the antiferromagnetic Heisenberg model,
\begin{equation}\label{eq:h}
\hat{\mathcal{H}} = J \sum_{\langle i,j \rangle} \hat{\mathbf{S}}_{i} \cdot \hat{\mathbf{S}}_{j} + \sum_{i} h_{i} \hat{S}^{z}_{i},
\end{equation}
where $i$ and $j$ denote nearest neighbors on a two-dimensional square lattice of size $L_1\times L_2$, and $ \hat{\mathbf{S}}_{i} =(\hat{S}^{x}_{i},\hat{S}^{y}_{i},\hat{S}^{z}_{i})^\text{T}$ is a quantum-mechanical spin-$\frac{1}{2}$ degree of freedom. The on-site magnetic fields $h_i$ are drawn randomly from a uniform distribution $[-W,W]$ in the case of quenched disorder, and all results are averaged over 1000 different disorder realizations. In the case of quasiperiodic disorder, we set $h_i=\Delta\cos(2\pi\beta_1x_i+\phi_1)+\Delta\cos(2\pi\beta_2y_i+\phi_2)$ where $x_i$ and $y_i$ denote the position in the directions associated with $L_1$ and $L_2$, respectively. We choose $\beta_1=0.721$ and $\beta_2=0.693$ \cite{Bordia2017,Kennes2018}, $\Delta$ is the disorder strength, and we average over 5000 realizations of the phases $\phi_{1,2}$. From now on, we pick $J=1$ as the unit of energy. In the vertical direction associated with $L_2$, we employ either open boundary conditions (OBC) or cylindrical boundary conditions (CBC); the system always has open boundaries in the horizontal direction associated with $L_1$. Our setup is illustrated in Fig.~\ref{fig:system}.

We determine a random excited eigenstate -- which is generically a high-energy, mid-spectrum state -- of the Hamiltonian in \eqref{eq:h} using the density-matrix renormalization group (DMRG). The DMRG is an accurate, tensor-network based method to study quantum systems \cite{White1992,White1993,Schollwoeck2005,Schollwoeck2011} which was originally devised to compute ground states in one dimension but was later extended to tackle two-dimensional problems \cite{Stoudenmire2012,Hyatt2020} as well as excited states in problems with disorder \cite{Khemani2016,Sheng2016,Kennes2016,Yu2017}, called DMRG-X. For small system sizes, the success of the DMRG-X in providing generic excited states was explicitly benchmarked against exact diagonalization results and then applied to access (what are believed to be generic) excited states of larger one-dimensional many-body localized systems. In this Letter, we combine these developments and, for the first time, investigate high-energy states in a two-dimensional system using the DMRG-X technique \cite{Khemani2016}.

The key idea of the DMRG-X is to start out in a random product state that is an eigenstate of $\hat{\mathcal{H}}$ in the limit of large~$h_i$. This state is trivially expressed as a matrix-product state $|\Psi\rangle = \sum_{\{\sigma_i\}} \text{tr}(A^{\sigma_1}\cdots A^{\sigma_L})|\sigma_1\ldots\sigma_L\rangle$, where $\sigma_i$ denotes a local basis of $\hat S^z_i$. The matrices $A^{\sigma_i}$ are then updated successively from left to right (and vice versa) by DMRG sweeps. In contrast to a variational ground state calculation, we do not choose the lowest-energy state during each update but pick the eigenstate which has maximum overlap with the prior one \cite{Khemani2016}. This accounts for the fact that localized eigenstates which have similar energy differ vastly in their spatial structures. The size $\chi$ of the matrices $A^{\sigma_i}$ is called the local bond dimension; it is a measure of the amount of entanglement in the system and is the key numerical control parameter. We successively increase $\chi$ after each 20 sweeps (see the left panel of Fig.~\ref{fig:dmrg}). The calculation is stopped whenever the rescaled energy variance has sufficiently converged. By implementing the DMRG-X approach using $U(1)$-symmetries, we can reach a maximum value of $\chi=64$. This allows us to obtain eigenstates with an energy variance close to machine precision for almost all parameter sets $\{L_1,L_2,W\}$ shown in this Letter (see the right panel of Fig.~\ref{fig:dmrg}).

\begin{figure}[t]
\centering
\includegraphics[width=0.95\linewidth]{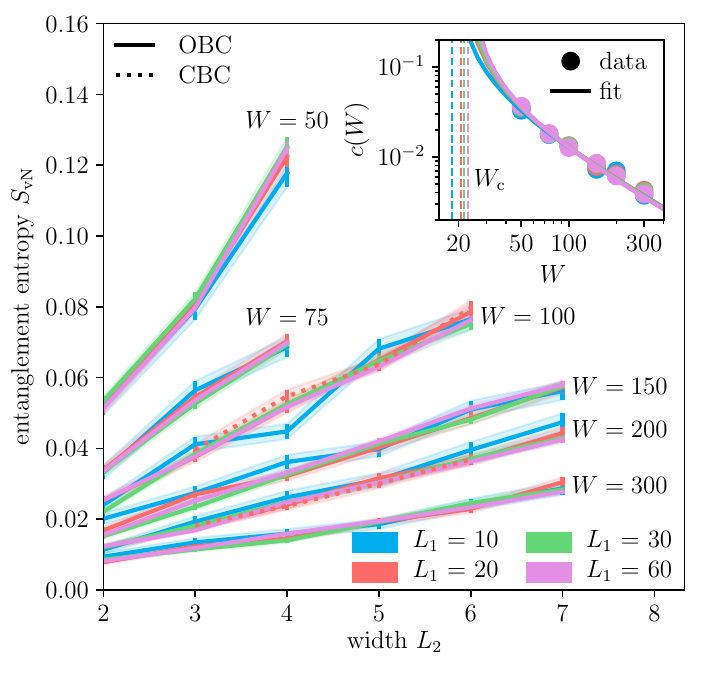}
\caption{The bipartite entanglement entropy of high-energy eigenstates of a lattice of length $L_1$ and width $L_2$ is computed using tensor networks for various strengths $W$ of quenched disorder and different boundary conditions. The system is cut vertically around the middle (see Fig.~\ref{fig:ent2} for details), and $S_{\text{vN}}$ grows linearly with $L_2$. This is compatible with an entanglement area law. The vertical lines indicate the error associated with the disorder averaging. \textit{Inset:} Proportionality constant $c(W)$ in $S_{\text{vN}}=c(W)L_2$. A divergence at a critical value $W_\text{ c}$ signals the crossover into an ergodic regime. We observe that $W_\text{c}$ shows a small, monotonous drift to larger values as $L_1$ increases (see the main text for further comments).}
\label{fig:ent}
\end{figure}

In order to apply the DMRG-X, our two-dimensional lattice is first transformed into a one-dimensional chain in a snakelike fashion (see Fig.~\ref{fig:system}). Due to this mapping, the Hamiltonian contains long-range terms. We choose a scheme where the snake connects the bottom of each column to the top of the next column, which has the distinct advantage that the local bond dimension of the matrix product operator that represents $\hat{\mathcal{H}}$, which scales linearly with $L_2$, is reduced compared to the standard approach where subsequent columns are traversed in the opposite order.

\section{Results}

\subsection{Entanglement entropy and quenched disorder}

We first study the von Neumann entropy,
\begin{equation}
S_\text{vN}=-\text{tr}\hat\rho\log_2\hat\rho, 
\end{equation}
where $\hat\rho$ is the reduced density matrix of one half of a system that is cut vertically. This is a measure of the {\it bipartite entanglement}. In a thermal ensemble, $S_\text{vN}$ is an extensive quantity that should scale with the volume $L_1L_2$ of the system. Due to the eigenstate thermalization hypothesis, the same should hold true for an eigenstate the local properties of which are reflective of the thermal ones at the corresponding energy. The behavior of a localized state, however, is very different: the area law stipulates that $S_\text{vN}$ is independent of $L_1$ but increases linearly with $L_2$, i.e., with the area between both halves of the bipartition. The entanglement entropy can thus be used to differentiate between those two phases of matter.

\begin{figure}[t]
\centering
\includegraphics[width=0.95\linewidth]{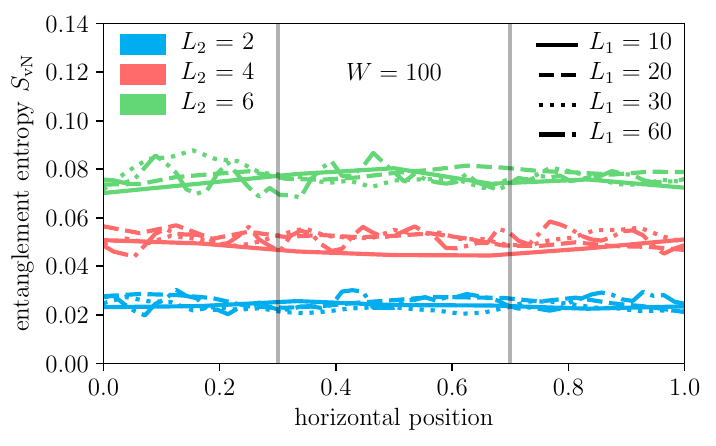}
\caption{Bipartite entanglement entropy as a function of the relative horizontal position where the lattice is cut vertically for various values of $L_1$ and $L_2$ (quenched disorder, OBC). The area law stipulates that $S_{\text{vN}}\sim L_2$, while being independent of $L_1$. We show a running average over three neighboring data points. The gray lines indicate the \enquote{middle area,} the average value of which is plotted in Fig.~\ref{fig:ent}. }
\label{fig:ent2}
\end{figure}

We begin with the case of quenched disorder. In Fig.~\ref{fig:ent}, we show $S_\text{vN}$ as a function of $L_2$ for various $L_1$, different disorder strengths $W$, and both open and cylindrical boundary conditions (note that CBC data are only included for $L_1=20$, $3\leq L_2\leq6$, $W\in\{75,100\}$). We averaged over all vertical cuts in the middle area of the system, which we defined as $\pm20\%$ around the center. One observes that the entanglement entropy does at most depend very slightly on $L_1$ but scales linearly with $L_2$. This points towards an area law and the existence of a many-body localized regime of the finite two-dimensional Heisenberg model at sufficiently large $W$ \footnote{We allow for a maximum bond dimension of $\chi=64$. While there is a general relationship between $\chi$ and $S_\text{vN}$, this relationship is not completely quantitative; e.g., different states with the same $\chi$ can yield different $S_\text{vN}$. The curves in Fig.~\ref{fig:ent} terminate whenever $\chi=64$ is not sufficient to reach convergence.}.

In Fig.~\ref{fig:ent2}, we show the entanglement entropy as a function of the relative position where the lattice is cut vertically for $W=100$ and different values of $L_1$ and $L_2$. The data points are averaged over three neighboring horizontal sites; this is meant as a visual aid to suppress noise related to the finite disorder sampling. These raw data again indicate an area law.

\begin{figure}[t]
\centering
\includegraphics[width=1\linewidth]{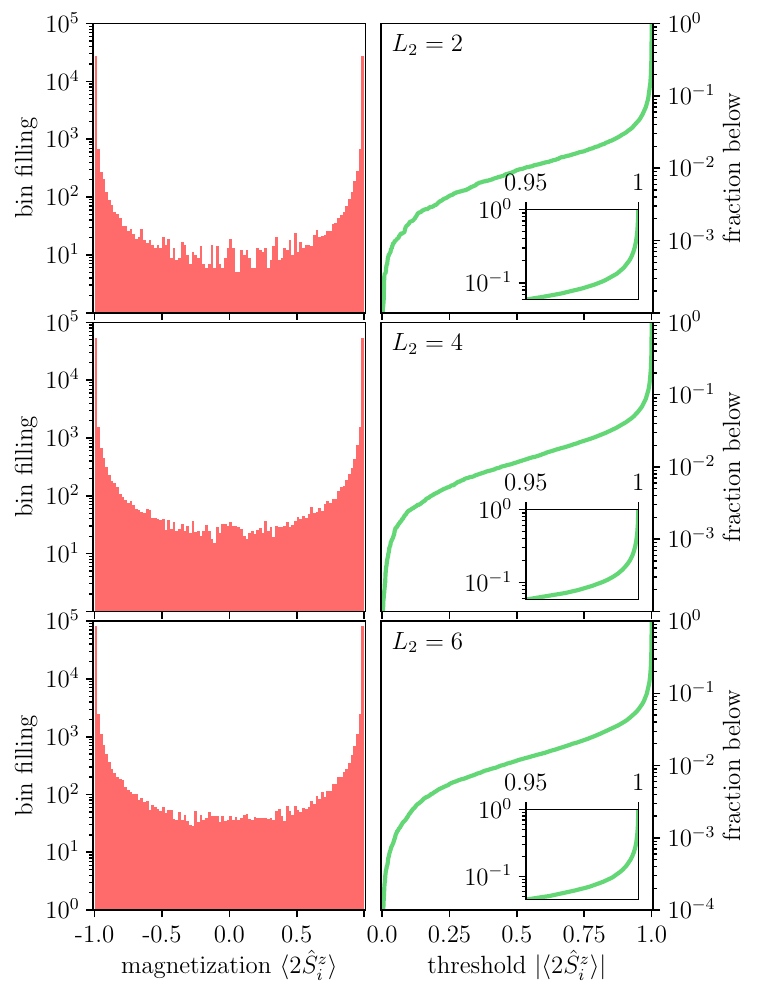}
\caption{Distribution of the magnetization $\langle 2\hat S^z_i\rangle$ in high-energy eigenstates for $L_1=30$, different $L_2$, and $W=100$ (quenched disorder). Due to the lack of ergodicity, the eigenstate thermalization hypothesis is violated. We do not observe a peak at $\langle 2\hat S^z_i\rangle=0$ but a strong pinning around $\langle 2\hat S^z_i\rangle=\pm1$. This is illustrated more clearly by plotting the fraction of values $|\langle 2\hat S^z_i\rangle|$ that lie below a given threshold.}
\label{fig:eth}
\end{figure}

As the strength of the disorder decreases, the entanglement entropy grows, and eventually the local bond dimension and thus the numerical effort become prohibitively large. This can be quantified by studying the prefactor $c(W)$ appearing in the area law, $S_\text{vN}=c(W)L_2$, which is plotted in the inset to Fig.~\ref{fig:ent}. The raw data is fitted to the functional form $c(W) = a / (W-W_\text{c})$. From this analysis, one estimates that the transition into the ergodic phase occurs at around $W_\text{c}\approx20$. This value of $W_\text{c}$ is in qualitative agreement with predictions from $\ell$-bit studies of the same system \cite{Chertkov2021}; evidence for MBL is also reported in \cite{Tang2021}. Importantly, this transition point is extracted directly from the area law scaling of the entanglement, a prototypical feature of many-body localization. One should note that $W_\text{c}$ increases monotonically with $L_1$; it is, however, unclear if this should be interpreted as evidence for an instability of the MBL phase in the thermodynamic limit.

\begin{figure}[t]
\centering
\includegraphics[width=0.9\linewidth]{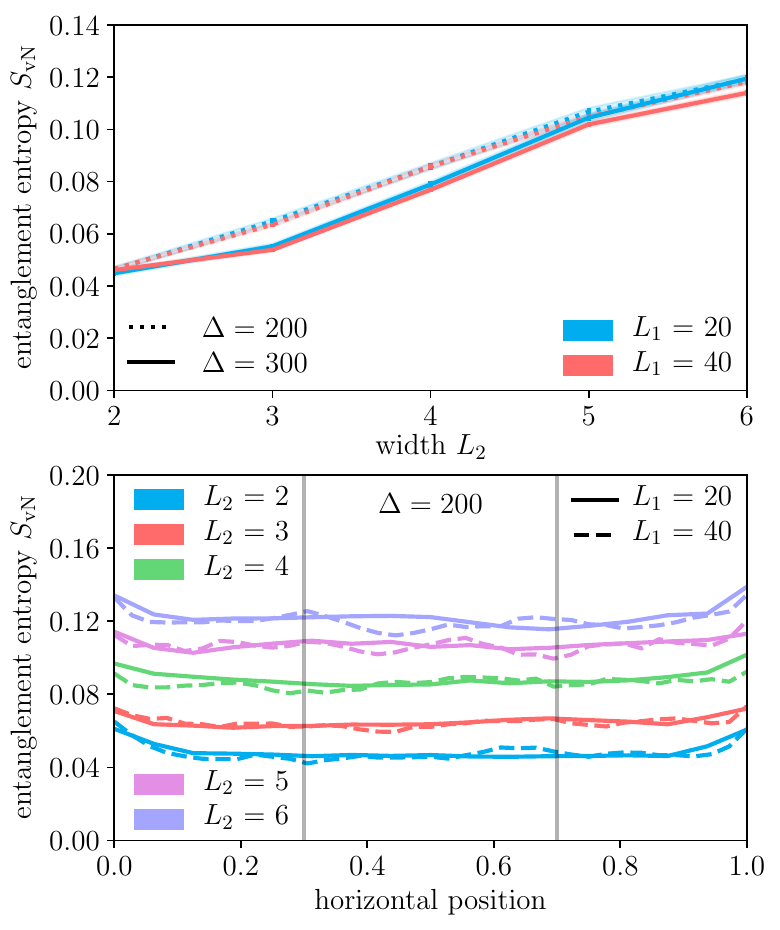}
\caption{The same as in Figs.~\ref{fig:ent} and \ref{fig:ent2} but for the case of quasiperiodic disorder $\Delta$ and OBC.}
\label{fig:quasi}
\end{figure}

\subsection{Eigenstate thermalization hypothesis}

Another hallmark of the localized phase is the lack of ergodicity. In an ergodic system, the eigenstate thermalization hypothesis dictates that local properties of generic eigenstates mirror the behavior of a thermal ensemble. In the left column of Fig.~\ref{fig:eth}, we plot the distribution of the local magnetization $\langle2\hat S^z_i\rangle$ for all sites $i$, fixed $L_1$, $W$ (quenched disorder), and various $L_2$. We do not observe a thermal peak at $\langle2\hat S^z_i\rangle=0$ but a strong pinning around $\langle2\hat S^z_i\rangle=\pm1$. Overall, the distribution is symmetric around $\langle2\hat S^z_i\rangle=0$. This illustrates that the eigenstate thermalization hypothesis is violated in these finite systems.

To quantify the strength of the pinning to $\langle2\hat S^z_i\rangle=\pm1$ and thus the non-ergodicity in the MBL regime more clearly, we plot in the right column of Fig.~\ref{fig:eth} the fraction of the spectrum of  $|\langle2\hat S^z_i\rangle|$ that lies below a given threshold. Varying the threshold from $0$ to $1$ (which correspond to the middle and the ends of the spectrum of $\langle2\hat S^z_i\rangle$, respectively) we find that only extremely little weight (less than 10\%) does not lie within a window of 5\% around the ends of the spectrum. Furthermore, the \mbox{growth of weight as one approaches the end of the spectrum} is faster than exponential, underlining the extreme pinning of the expectation values of the local spins to  $\langle2\hat S^z_i\rangle=\pm 1$.

\subsection{Quasiperiodic disorder}

Next, we turn to the case of quasiperiodic disorder. The results are summarized in Fig.~\ref{fig:quasi}, which is analogous to Figs.~\ref{fig:ent} and \ref{fig:ent2}. While the data hint towards an area law for large disorder strength $\Delta$ (see also \cite{arxiv.2108.08268,arxiv.2204.05198}), it was not feasible numerically to address the transition into the ergodic phase. We observe that a comparable disorder strength leads to a significantly larger entanglement entropy.

\section{Discussion}

Our findings provide further evidence for many-body localization of systems in higher than one dimension. In particular, we studied the behavior of the entanglement entropy as the two-dimensional limit is approached and found an area law scaling as $L_2$ is increased at fixed $L_1>L_2$. We employed an accurate tensor-network based method that circumvents most of the criticism applicable to other approaches such as weak coupling expansions or approximate effective representations like the $\ell$-bit picture, which are very powerful but hard to converge in two dimensions \cite{Chertkov2021}. It is often argued that ultimately, many-body localization is unstable in higher than one dimension, and we believe that our data can contribute to this discussion.

The area law in localized systems renders them an ideal playground for a tensor-network approach, the accuracy and complexity of which grow with the amount of entanglement in the system. However, the numerical effort is still substantial, since the complexity to capture an arealaw in two dimensions roughly corresponds to the one of a volume law in one dimension. On the other hand, the amount of entanglement decreases with the strength of the disorder, and a \enquote{sweet spot} can be found trading disorder strength for system size in the quenched case. By exploring the scaling with respect to system size at this sweet spot, we can extrapolate a finite critical disorder strength where the transition to the ergodic phase occurs.

Future investigations could point out similarities and possibly differences of many-body localization in varying dimensions, e.g., the transport behavior and the perspective of MBL in two dimensions for the imminent age of noisy intermediate-scale quantum computing \cite{Preskill2018quantumcomputingin,Kshetrimayum2021}. Finally, if two dimensions are not enough to suppress the presence of a many-body localized regime, it begs the question: Is there a finite dimensionality $d_c$ at which MBL indeed will be unstable?

\section*{Acknowledgements}

K.D.~and C.K.~acknowledge support by the Deutsche Forschungsgemeinschaft (DFG) through the Emmy Noether program (Grant No.~KA3360/2-1) as well as by \enquote{Nieders\"achsisches Vorab} through the \enquote{Quantum- and Nano-Metrology (QUANOMET)} initiative  within  the Project No.~P-1. D.M.K.~was supported by the DFG via  Germany’s Excellence Strategy -- Cluster of Excellence Matter and Light for Quantum Computing (ML4Q, Project No.~EXC 2004/1, Grant No.~390534769). We acknowledge support from the Max Planck-New York City Center for Non-Equilibrium Quantum Phenomena.

\bibliography{references}

\end{document}